\documentclass[10pt,aps,prx,showpacs,superscriptaddress,preprint,citeautoscript]{revtex4-1}

\usepackage{amsmath}
\usepackage{amsfonts}
\usepackage{amssymb}
\usepackage{bm}
\usepackage{graphicx}
\usepackage{color}
\usepackage{array}

\usepackage{hyperref}

\usepackage{tikz}

\def\beq#1#2\eeq{\begin{equation}\label{#1}#2\end{equation}}
\def\bal#1#2\eal{\begin{align}\label{#1}#2\end{align}}
\def\bse#1#2\ese{\begin{subequations}\label{#1}#2\end{subequations}}
\def\ba{\begin{aligned}}
\def\ea{\end{aligned}}

\newcommand{\inc}{\mathrm{inc}}
\newcommand{\sca}{\mathrm{scat}}
\newcommand{\cyl}{\mathrm{cyl}}
\newcommand{\diff}[2]{\frac{\partial{#1}}{\partial{#2}}}
\newcommand{\mat}{\left[\begin{matrix}}
\newcommand{\emat}{\end{matrix}\right]}
\newcommand{\ph}{\mathrm{ph}}
\newcommand{\gr}{\mathrm{gr}}
\newcommand{\eps}{\varepsilon}

\def\dd{\operatorname{d}}

\def\diag{\operatorname{diag}}

\begin{document} %%%%%%%%%%%%%%%%%%%%%%%%%%%%%%%%%%%%%%%%%%%%%%%%%%%%%%%%%%%%%%%%%%%%%%%%%%%%%%%%%%%%%%%%%%%

\title[Willis cylinder scattering]{Acoustic scattering from a fluid cylinder with Willis constitutive properties}
\author{Michael B. Muhlestein}
\affiliation{U.\ S.\ Army Engineer Research and Development Center, 72 Lyme Rd., Hanover, NH 03755}
\author{Benjamin M. Goldsberry}
\affiliation{Department of Mechanical Engineering and Applied Research Laboratories, The University of Texas at Austin, 10000 Burnet Rd., Austin, TX 78758}
\author{Andrew N. Norris}%
\affiliation{Department of  Mechanical and Aerospace Engineering, Rutgers University, 
Piscataway, NJ 08854}
\author{Michael R. Haberman}
\affiliation{Department of Mechanical Engineering and Applied Research Laboratories, The University of Texas at Austin, 10000 Burnet Rd., Austin, TX 78758}

\begin{abstract}
	A material that exhibits Willis coupling has constitutive equations that couple the pressure-strain and momentum-velocity relationships.  This coupling arises from subwavelength asymmetry and non-locality in heterogeneous media.  This paper considers the problem of the scattering of a plane wave by a cylinder exhibiting Willis coupling using both analytical and numerical approaches.  First, a perturbation method is used to describe the influence of Willis coupling on the scattered field to a first-order approximation.  A higher-order analysis of the scattering based on generalized impedances is then derived.  Finally, a finite element method-based numerical scheme for calculating the the scattered field is presented.  These three analyses are compared and show strong agreement for low to moderate levels of Willis coupling.
	
\end{abstract}

\maketitle

\section{Introduction}

Recent homogenization research relevant to the topic of metamaterials has noted that acoustical systems with subwavelength asymmetry in properties or structure cannot be adequately described in terms of the standard material properties: mass density and bulk modulus.\cite{alu2011,sieck2015,muhlestein2016a}  These and other similar systems may be described more appropriately with the Willis constitutive equations which couple the acoustic pressure and the momentum density to both the volume strain and the particle velocity using an additional material property called the Willis coupling vector.  This additional material property is analogous to bianisotropy in electromagnetism,\cite{sieck2015} and is attractive to designers of acoustic metamaterials as it opens a new dimension of material parameter space relative to standard materials.  One potential application of Willis materials, or materials with non-trivial Willis coupling vectors, uses scattering for localization, imaging, and classification of objects.

Scattering of mechanical waves from Willis materials has received only limited and tangential attention.  Muhlestein and Haberman\cite{muhlestein2016} used a Green's function approach to describe the total displacement field in a Willis elastic matrix with Willis inclusions in the long-wavelength limit, but restricted their analysis to the field immediately surrounding the inclusions.  On the other hand, electromagnetic scattering from bi-anisotropic materials has received more attention.  Lakhtakia used a Green's function-based approach to describe Rayleigh (long-wavelength) scattering from bi-anisotropic ellipsoids within a bi-isotropic background material,\cite{lakhtakia1991} and described scattering from more general geometries using a discrete-dipole approximation.\cite{lakhtakia1992}  Jakoby used a propagator matrix formalism to describe scattering of obliquely incident electromagnetic plane waves from impedance cylinders with inhomogeneous bi-anisotropic coatings.\cite{jakoby1997}  Zhang, \emph{et al}.\ studied the scattering from arbitrary three-dimensional bi-anisotropic materials using a hybrid finite element-boundary integral method.\cite{zhang2004}  The problem considered here may be considered as a generalization of previous studies of scalar wave scattering from circular domains in the context of acoustics with anisotropic density,\cite{torrent2009} anti-plane (SH) elastic waves with anisotropic stiffness,\cite{bostrom2015} and two-dimensional electromagnetics with anisotropic permitivity and permeability.\cite{wu1995}

The purpose of this paper is to provide an analytical foundation for scattering of acoustic plane waves from Willis-fluid cylinders.  The basic equations of a Willis material are introduced in Sec.~\ref{sec1} which also describes its anisotropic wave equation.  In Sec.~\ref{sec2} an exact analysis of two-dimensional scattering (normal incidence on infinite cylinders) is provided.  Since the resulting equations that describe the scattered field do not have analytical solutions, two types of asymptotic expansions for weak Willis coupling are used to provide an explicit description of the scattered field.  Section~\ref{sec:numeric} describes a finite element-based approach to the same problem, which is then compared with and validates the analytical predictions.  Some final thoughts are then provided in Sec.~\ref{sec:conclusions}.

\section{Willis Materials}\label{sec1}

A Willis fluid may be described by the constitutive equations
\begin{subequations}\label{eq:origConst}
\begin{align}
- p & = \kappa \varepsilon +   \vec\psi\cdot \dot{\vec v}, \label{eq:pressure} \\
\vec\mu & = \bm\rho\cdot \vec v +  \vec\psi \dot\eps, \label{eq:momentum}
\end{align}
\end{subequations}
where $p$ is the acoustic pressure (hereafter just pressure), $\eps$ is the volume strain, $\vec v$ is the particle velocity (hereafter just velocity), $\vec\mu$ is the momentum density, $\kappa$ is the bulk modulus, $\bm\rho$ is the effective mass density tensor, and $\vec\psi$ is the Willis coupling vector.  For this analysis, the material properties are assumed to be constants with respect to frequency.  Note that the standard constitutive equations are recovered if $\vec\psi\rightarrow0$ and $\bm\rho\rightarrow\rho\bm I$ where $\rho$ is the standard mass density scalar and $\bm I$ is the second-order identity tensor.\cite{willis1981,willis1997}  Assuming a time-harmonic acoustic field ($e^{-i\omega t}$ time convention), the constitutive equations may be inverted to yield
\begin{subequations}
\begin{align}
	\varepsilon &= \Delta^{-1}\left[ \frac{-p}{\kappa} + \frac{i\omega}{\kappa}\vec\psi\cdot\bm\rho^{-1}\cdot\vec\mu \right], & \Delta &= 1 + \frac{\omega^2}{\kappa}\vec\psi\cdot\bm\rho^{-1}\cdot\vec\psi, \\
	\vec v &= \bm\Delta^{-1}\cdot\left[ \frac{-i\omega}{\kappa}\bm\rho^{-1}\cdot\vec\psi p + \bm\rho^{-1}\cdot\vec\mu \right], &	
	\bm\Delta &= \bm I + \frac{\omega^2}{\kappa}\left(\bm\rho^{-1}\cdot\vec\psi\right)\otimes\vec\psi,
\end{align}
\end{subequations}
where $\otimes$ is the tensor product.  For simplicity of this initial analysis, only isotropic mass density tensors will be considered such that $\bm\rho=\rho\bm I$.  Then the inverted constitutive equations simplify to
\begin{subequations}
\begin{align}
	\varepsilon &= \Delta^{-1}\left[ \frac{-p}{\kappa} + \frac{i\omega}{\rho\kappa}\vec\psi\cdot\vec\mu \right], & \Delta &= 1 + \frac{\omega^2}{\rho\kappa}\vec\psi\cdot\vec\psi, \\
	\vec v &= \bm\Delta^{-1}\cdot\left[ \frac{-i\omega}{\rho\kappa}\vec\psi p + \frac{\vec\mu}{\rho} \right], &	
	\bm\Delta &= \bm I + \frac{\omega^2}{\rho\kappa}\vec\psi\otimes\vec\psi.
\end{align}
\end{subequations}
The constitutive equations may be further simplified with the definition of the non-dimensional ``asymmetry factor''
\begin{equation}
\vec W =  \frac{\omega \vec\psi}{Z}
\end{equation}
and the wavenumber $k=\omega/c$, where $Z=\rho c$ is the characteristic impedance and $c=\sqrt{\kappa/\rho}$ is the wave speed.  When combined with the dynamic equation $\dot{\vec\mu}=-\nabla p$,  the constitutive equations may then be written as
\bse{5}
\bal{5a}
\varepsilon & = \frac {-1}{\omega Z} (1+ W^2)^{-1} \big( k p - \vec W\cdot \nabla p \big) ,
\\
\vec v &= \frac {-i}{\omega \rho } \big( \bm I + \vec W\otimes \vec W  \big)^{-1} \cdot \big( k \vec W p +\nabla p  \big) ,
\label{5b}
\eal
\ese
where $W^2 = \vec W\cdot\vec W$.  Note that Eq.~\eqref{5b} may also be written as
\begin{equation}\label{simpVel}
	\vec v = \frac{-i}{\omega\rho}(1+W^2)\left( kp\vec W - (\vec W\cdot\nabla p)\vec W + (1+W^2)\nabla p \right),
\end{equation}
which may be verified by multiplication by $\bm I+\vec W\otimes\vec W$.

Using the definition of the volume strain rate $\dot\varepsilon=\nabla\cdot\vec v$, 
Eqs.~\eqref{5a} and \eqref{simpVel} may be combined into a single anisotropic wave equation for the acoustic pressure, 
\beq{6}
(1+W^2) \nabla^2  p - (\vec W\cdot \nabla)^2 p + k^2 p = 0.
\eeq 
In order to see the anisotropy in detail, let $p = p_0 e^{i \vec\xi\cdot {\vec x}}$, with vector wavenumber $\vec\xi = \xi \hat n$ and $|\hat n|=1$.  Then Eq.~\eqref{6} may be written as
\beq{6.1-}
\left(1+W^2 - (\vec W\cdot \hat n)^2 \right) \xi^2 = k^2.
\eeq
%\begin{align}
%	-\xi^2(1+W^2)\left[ \bm I + \vec W\otimes\vec W \right]^{-1}:(\hat n\otimes\hat n)\ +\ & \notag \\
%	i\xi k\left\{ (1+W^2)\left[ \bm I+\vec W\otimes\vec W \right]^{-1}:(\vec W\otimes\hat n) - \vec W\cdot\hat n \right\} + k^2p &= 0.
%\end{align}
This provides an equation for the phase velocity $c_\ph=\omega/\xi$ in the $\hat n$ direction:
\beq{6.1}
c_\ph^2 = \left(1+W^2 - (\vec W\cdot \hat n)^2 \right)  \, c^2 .
\eeq
Notice that the phase velocity is a minimum and equal to $c$ for $\hat n = \pm\vec W/W$, and is a maximum and equal to $c\sqrt{1+W^2}$ in the directions orthogonal to $ \vec W $.  The group velocity vector is defined as $\vec c_\gr = \frac{\dd \omega }{\dd \vec\xi} $. It may be evaluated as  $\vec c_\gr = \frac{1}{2\omega }\frac{\dd c^2_\ph\xi^2}{\dd \vec\xi}$ using Eq.~\eqref{6.1} for $c_\ph^2$, which  gives 
\beq{03}
\vec c_\gr = \big[ (1+W^2)\hat n - (\vec W\cdot \hat n) \vec W\big]c_\ph . 
\eeq
As is common in dealing with anisotropic wave equations, it is  instructive to consider the inverse of the phase speed, i.e.\ the slowness $s= 1/c_\ph$.  The slowness surface, i.e. the surface $S(\vec x) = s\hat n$, therefore has the form of a prolate spheroid.  It may be checked that the direction of the group velocity is perpendicular to the slowness surface, a result that is standard in anisotropic elastic solids.\cite{musgrave2003}  Note that Eq.~\eqref{03} takes into account the assumption that $\vec W$ is a linear function of $\omega$.  If $\vec W$ is independent of $\omega$ the group velocity becomes  
\beq{0=3}
\vec c_{\gr0} = \frac{c^2}{c_\ph}\, \big[ (1+W^2)\hat n - (\vec W\cdot \hat n) \vec W\big] .    
\eeq
 Note that $\vec c_{\gr0} \cdot \hat n = c_\ph$ which is typical of frequency independent anisotropy,\cite{musgrave2003} whereas $\vec c_{\gr} \cdot \hat n = c_\ph^3/c^2$.  In summary, the dependence of $\vec W$ on $\omega$ leaves the direction of $\vec c_\gr$ unchanged while its magnitude is increased by a factor $\frac{c_\ph^2}{c^2} \ge 1$.

\section{Willis Cylinder Scattering}\label{sec2}

\begin{figure}
	\centering
	\includegraphics[width=3.315in]{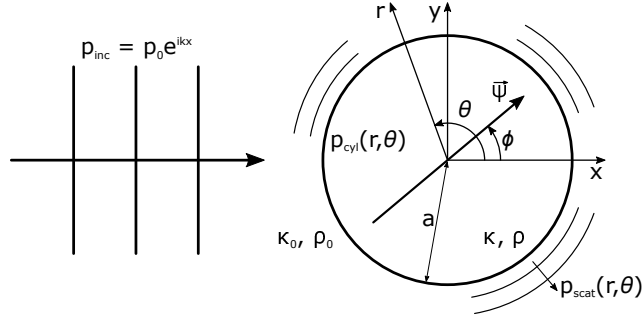}
%	\begin{tikzpicture}[scale=0.8]
%	\draw[very thick] (0,0) circle (2);
%	\draw[<->] (0,2.5) node[left]{$y$} -- (0,0) -- (2.5,0) node[below]{$x$};
%	\draw[->,thick] (-1.5,-1) -- (1.5,1) node[left=2mm]{\color{black}$\vec\psi$};
%	\draw (1,0) arc (0:33.69:1) node[pos=0.5,right]{$\phi$};
%	\draw[->] (0,0) -- (-1,2.5) node[left=2mm]{\color{black}$r$};
%	\draw (0.7,0) arc (0:180-68.20:0.7) node[pos=0.6,above]{$\theta$};
%	\draw[<->,thin] (-0.008,-0.049) -- (-0.305,-1.926) node[pos=0.5,left]{$a$};
%	\draw (-3.5,-1.5) -- (-3.5,1.5);
%	\draw (-4.5,-1.5) -- (-4.5,1.5) node[above=1]{$p_\inc=p_0e^{ikx}$};
%	\draw (-5.5,-1.5) -- (-5.5,1.5);
%	\draw[->] (-6,0) -- (-3,0);
%	\draw[domain=20:60] plot ({2.3*cos(\x)},{2.3*sin(\x)}); \draw[domain=20:60] plot ({2.5*cos(\x)},{2.5*sin(\x)});
%	\draw[domain=-30:-80] plot ({2.3*cos(\x)},{2.3*sin(\x)}); \draw[domain=-30:-80] plot ({2.5*cos(\x)},{2.5*sin(\x)});
%	\draw[domain=140:170] plot ({2.3*cos(\x)},{2.3*sin(\x)}); \draw[domain=140:170] plot ({2.5*cos(\x)},{2.5*sin(\x)});
%	\draw[->] (1.205,-1.720) -- (1.549,-2.212) node[right]{$p_\sca(r,\theta)$};
%	\draw (-1.05,0.25) node{$p_\cyl(r,\theta)$};
%	\draw (0.75,-1) node{$\kappa,\rho$};
%	\draw (-2.25,-1.75) node{$\kappa_0,\rho_0$};
%	\end{tikzpicture}
	\caption{\label{schematic}Schematic of a plane wave scattering from a Willis-coupled cylinder of radius $a$.  The incident wave has a wavenumber $\vec k_0=k_0\hat x$.  The background material has mass density $\rho_0$ and bulk modulus $\kappa_0$ and the cylinder has mass density $\rho$, bulk modulus $\kappa$, and coupling vector $\vec\psi$.}
\end{figure}

Consider a Willis cylinder of radius $a$, bulk modulus, mass density, and Willis coupling vector of $\kappa$, $\rho$, and $\vec\psi$, respectively, in a background fluid with bulk modulus $\kappa_0$,  mass density  $\rho_0$, impedance $Z_0=\sqrt{\rho_0\kappa_0}$, and wavenumber $k_0=\omega\sqrt{\rho_0/\kappa_0}$.  Let the origin of a Cartesian coordinate system be set in the center of the cylinder with the $z$-axis parallel to the cylinder axis.  A schematic of this situation is shown in Fig.~\ref{schematic}.  It is worth noting that the cylinder is assumed to have no boundary layer, meaning that the material properties of the cylinder are homogeneous throughout the entire cylinder including right at the edges.  This assumption is equivalent to assuming that the microstructure is sufficiently small compared to a wavelength that interface effects are negligible.\cite{srivastava2017}

Now, consider an incident plane wave propagating in the $x$ direction.  The incident wave may be written as
\begin{equation}\label{123}
	p_\inc = p_0e^{ikx} = p_0\sum_{m=-\infty}^\infty i^mJ_m(k_0 r)e^{im\theta},
\end{equation}
where $p_0$ is the pressure amplitude and $J_m$ is the $m^\text{th}$ Bessel function of the first kind.  The scattered pressure field may be written as
\begin{align}
	p_\sca &= p_0\sum_{m=-\infty}^\infty A_mH_m^{(1)}(k_0r)e^{im\theta}, \label{eq:psca}
\end{align}
where $H_m^{(1)}$ is the $m^\text{th}$ order Hankel function of the first kind.  The pressure inside the cylinder satisfies Eq.~\eqref{6} and can be converted into isotropic form by rescaling the coordinates, which allows the separation of variables solution:
\beq{7}
p_\cyl = \sum_n C_n J_n (kR) e^{i n \gamma}
\ \ \text{where } \ \ 
\ba R&= r \sqrt{\frac{1 + W^2 \cos^2(\theta - \phi)}{1 + W^2} }, 
\\ 
\gamma &= \tan^{-1} \frac{\tan (\theta - \phi)}{\sqrt{1 + W^2} } ,
\ea 
\eeq 
and where $\phi$ denotes the direction of $\vec W$ in terms of the regular polar coordinates $r,\theta$, that is $\vec W = W \hat r(\phi)$ where $ \hat r(\theta) = \vec r/r$  is the unit vector in the radial direction. 

The boundary conditions are continuity of pressure and normal component of the velocity at the surface of the cylinder. The latter follows from Eq.~\eqref{5b} as
\beq{8}
v_r \equiv {\vec v}\cdot \hat r
 = \frac {-i}{\rho \omega} \Big( 
\partial_r p  + \frac{\vec W\cdot \hat r}{1+W^2}
\big( kp - \vec W\cdot \nabla p\big) 
\Big)
\eeq 
Equations~\eqref{7} and \eqref{8}  provide an exact solution inside the circular cylindrical  scatterer of radius $a$. 
The difficulty arises in trying to match the interior solution to the exterior one.  Specifically, the representation of $p$ in Eq.~\eqref{7} does not translate to a simple one in terms of $r,\theta$.  We therefore assume  $W \ll1$ and consider asymptotic expansions of the solution in terms the small coupling parameter $W$. 
	
\subsection{First Order Approximation} \label{sec2B} %%%%%%%%%%%%%%%%%%%%%%%%%%%%%%%%%%%%%%%
At this level of approximation we consider only the contributions of order $W$ in the equation for the velocity in Eq.~\eqref{simpVel} and the pressure in Eq.~\eqref{6}, which become, respectively, 
\begin{equation}
	{\vec v} = -i\frac{\vec W}{Z}p - \frac{\nabla p}{kZ} \hspace{15mm}\text{and}\hspace{15mm}\nabla^2p + k^2 p = 0,
\end{equation}
where $Z=\sqrt{\rho\kappa}$ is the characteristic impedance of the Willis material.  The solution in the cylinder is therefore 
\beq{77}
		p_\cyl = p_0\sum_{m=-\infty}^\infty B_mJ_m(kr)e^{im\theta},
\eeq
where, referring to Eq.~\eqref{7}, $B_m=C_m e^{-im\phi}$. 
 Continuity of the pressure combined with orthogonality yields
\begin{equation}\label{eq:Pbc}
	i^mJ_m(k_0a) + A_mH_m^{(1)}(k_0a) = B_mJ_m(ka).
\end{equation}
The condition for continuity of the normal component of velocity is more complicated to derive (see Appendix~\ref{appendix:velocity}), but results in
\begin{align}
	\frac{Z}{Z_0}&\left[i^mJ'_m(k_0a) + A_mH_m^{(1)'}(k_0a)\right] \notag \\
	&\hspace{5mm} = B_mJ'_m(ka) + \frac{W}{2}\left[e^{-i\phi}B_{m-1}J_{m-1}(ka) + e^{i\phi}B_{m+1}J_{m+1}(ka)\right]. \label{eq:Vbc}
\end{align}

Since Eq.~\eqref{eq:Vbc} depends on $B_{m-1}$, $B_m$, and $B_{m+1}$, it becomes impractical to determine $A_m$ and $B_m$ in closed form from Eqs.~\eqref{eq:Pbc} and \eqref{eq:Vbc}.  
A perturbation analysis, however, may be used to provide explicit expressions up to first order.  For $W\ll1$ but $\ne0$, the coefficients may be expanded in a series over $W$ as
\bse{1=3}
\begin{align}
	A_m &= W^0A_m^{(0)} + W^1A_m^{(1)} + W^2A_m^{(2)} + \cdots, \\
	B_m &= W^0B_m^{(0)} + W^1B_m^{(1)} + W^2B_m^{(2)} + \cdots. 
	\end{align}
\ese
Substituting these expansions into Eqs.~\eqref{eq:Pbc} and \eqref{eq:Vbc} and setting 
$W=0$ (no Willis coupling), the leading coefficients may be written as
 \bse{eq:am0}
\begin{align}
	%\left.A_m\right|_{W=0} \equiv 
	A_m^{(0)} &= -i^m\frac{Z J_m(ka)J'_m(k_0a) - Z_0 J'_m(ka)J_m(k_0a)}{Z J_m(ka)H^{(1)'}_m(k_0a) - Z_0J'_m(ka)H^{(1)}_m(k_0a)}, \\
	%\left.B_m\right|_{W=0} \equiv 
	B_m^{(0)} &= \frac{2i^{m+1} Z (\pi k_0a)^{-1}}{Z J_m(ka)H^{(1)'}_m(k_0a) - Z_0J'_m(ka)H^{(1)}_m(k_0a)},  
\end{align}
\ese
which is the classic result of scattering from a fluid cylinder.  
Substituting from Eq.~\eqref{1=3} into Eqs.~\eqref{eq:Pbc} and \eqref{eq:Vbc}, differentiating with respect to $W$ and setting $W=0$ yields 
\bse{3=9}
\begin{align}
	A_m^{(1)} &= \frac{Z_0}{2}\frac{e^{-i\phi}B_{m-1}^{(0)}J_m(ka)J_{m-1}(ka)+e^{i\phi}B_{m+1}^{(0)}J_m(ka)J_{m+1}(ka)}{Z J_m(ka)H^{(1)'}_m(k_0a) - Z_0J'_m(ka)H^{(1)}_m(k_0a)},   \\
	B_m^{(1)} &= \frac{Z_0}{2}\frac{e^{-i\phi}B_{m-1}^{(0)}H_m(k_0a)J_{m-1}(ka)+e^{i\phi}B_{m+1}^{(0)}H_m(k_0a)J_{m+1}(ka)}{Z J_m(ka)H^{(1)'}_m(k_0a) - Z_0J'_m(ka)H^{(1)}_m(k_0a)}.  
\end{align}
\ese
Combining Eqs.~\eqref{eq:psca} and \eqref{77} with \eqref{eq:am0} and \eqref{3=9} gives the first order approximation to the scattered and interior fields, 
\beq{0-2}
	A_m \approx A_m^{(0)} + WA_m^{(1)}, \ \  B_m \approx B_m^{(0)} + WB_m^{(1)}. 
\eeq

An important limiting case is that of $ka,k_0a\ll1$.  For mathematical concreteness, let $ka=\eta k_0a$ and assume $k_0a$ is small and $\eta$ is of order 1.  In this case one finds the $m=0,\pm1$ scattering coefficients dominate and may be approximated as
\begin{subequations}
\begin{align}
	A_0 & \approx (k_0a)^2\frac{\pi}{4}\left[\frac{Z-\eta Z_0}{iZ} + \frac{W\eta Z_0}{Z_0+\eta Z}2\cos(\phi)\right], \\
	A_{\pm 1} & \approx \pm (k_0a)^2 \frac{\pi}{4}\left[ \frac{Z_0-\eta Z}{Z_0+\eta Z} - \frac{iW\eta Z_0 }{Z_0+\eta Z}e^{\mp i\phi}\right].
\end{align}
\end{subequations}
Thus in the long wavelength limit the presence of a uniform Willis coupling modifies the relative strength of the monopole and dipole moments as a function of incidence angle.

\subsection{Higher Order Approximation}\label{sec3b}  %%%%%%%%%%%%%%%%%%%%%%%%%%%%%%%%%%%%%%%%%%%%%%%%%

As shown above, the traditional approach to determining the scattered acoustic fields becomes difficult for higher-order approximations of the pressure equation in Eq.~\eqref{6}.  However, a more general approach\cite{bobrovnitskii2006} to acoustic scattering that partitions the solution into three distinct physically meaningful {\it impedances} reduces the problem to the easier task of finding one of the impedances.  As before, assume that the total acoustic pressure $p$ comprises the incident, $p_\text{inc}$, and scattered, $p_\text{scat}$, components
\beq{b1}
p = p_\text{inc} + p_\text{scat}   
\eeq
which for the moment treated as vectors with an infinite number of components and will later be identified as the coefficients of $e^{im\theta}$.  The radial part of the velocity may be also be written as infinite vectors as
\beq{b2}
v_r = v_{r,\text{inc}} + v_{r,\text{scat}}.
\eeq
Surface impedance matrices $Z_\text{inc}$, $Z_\text{scat}$ and $Z_\text{cyl}$ are then introduced, which are defined such that on the surface bounding the scatterer from the exterior fluid  
\bse{b345}
\bal{b3}
p+ Z_\text{cyl}   v_r & = 0, 
\\
p_\text{inc} + Z_\text{inc} v_{r, \text{inc}} & = 0, \label{b4}
\\
p_\text{scat} - Z_\text{scat} v_{r,\text{scat}} & = 0. \label{b5}
\eal
\ese
Assuming that the impedances are known, the solution for the scattered field is just
\beq{b6}
p_\text{scat} = S\, p_\text{inc}, % \ \ v_{r,\text{scat}} = Q v_{r,\text{inc}}, 
\eeq
where the scattering matrix is 
\bal{b7}
%Q &= \big(Z_\text{scat}+Z_\text{cyl} \big)^{-1}\big(Z_\text{inc}-Z_\text{cyl} \big), \\
S &= \big(Y_\text{scat}+Y_\text{cyl} \big)^{-1}\big(Y_\text{inc}-Y_\text{cyl} \big),
%\label{b8}
\eal
and  $Y_\text{inc} = Z_\text{inc}^{-1}$, $Y_\text{scat}= Z_\text{scat}^{-1}$ and $Y_\text{cyl}= Z_\text{cyl}^{-1}$ are mobility matrices.  

In the case considered here the surface is circular, so that the infinite vectors $p$, $v_r$, etc. in Eq.~\eqref{b345} represent the components of these physical quantities in terms of $e^{in\theta}$ where $\theta $ is the polar angle and $n$ are integers.  We use the standard representation for the incident and scattered pressure,
\bal{1=1}
\big( p_\text{inc} , p_\text{scat} \big) 
&=   p_0 \sum_n   \big( F_n J_n(k_0r), A_n H_n^{(1)}(k_0r)\big) 
 \Big)  \, e^{i n \theta }
\notag \\
&=  p_0\sum_n  \Big( \hat{F}_n \frac{J_n(k_0r)}{J_n(k_0a)},  \hat{A}_n \frac{ H_n^{(1)}(k_0r)}{ H_n^{(1)}(k_0a)} \Big) \, e^{i n \theta }
\eal
where (see Eq.~\eqref{123}) $F_n = i^n$ for the assumed plane wave incidence.  Equation \eqref{b6} then becomes 
\beq{1=-1}
\hat{A} = S \hat{F}
\eeq
where $\hat{F}$ and $\hat{A}$ are vectors with elements $\hat{F}_n$, $\hat{A}_n$.  Alternatively, using the original $F$ and $A$ matrices we may write
\beq{1=-2}
A = \diag \big( 1/ H_n^{(1)}(k_0a) \big)  S    \diag \big( J_n(k_0a) \big)\, F,
\eeq
where $\diag(x_n)$ denotes a diagonal matrix with $x_n$ being the $(n,n)^\text{th}$ element.  Note that $Z_\text{inc}$ and $Z_\text{scat}$ (and hence $Y_\text{inc}$ and $Y_\text{scat}$) are diagonal with components
\bse{-5252}
\bal{-52}
[Z_\text{inc}]_{mn}  &= -i Z_0 \frac{J_n(k_0a)}{J_n'(k_0a)}\, \delta_{mn}, 
\\
[Z_\text{scat}]_{mn}  &= i Z_0 \frac{H_n^{(1)}(k_0a) }{H_n^{(1)'}(k_0a) } \, \delta_{mn}. 
\label{-53}
\eal
\ese
The main difficulty is with the impedance or mobility matrices for the cylinder itself. The total fields on the cylinder surface may be represented as 
\beq{15}
\big( v_r (a), p(a)\big) 
 =   \sum_n  \big( V_n, P_n\big)\, e^{i n \theta } , 
\eeq
and writing the elements of the cylinder admittance matrix $Y_\text{cyl}$ as $Y_{mn}$, the coefficients $V_m$ may be written as
\beq{16}
	V_m = \sum_n Y_{mn} P_n.
\eeq
Knowledge of $Y_\text{cyl}$ is crucial to evaluating the scattered field.  Methods for estimating $Y_\text{cyl}$ are discussed next.

\subsubsection{Perturbation solution}

A perturbation approach provides a useful means of approximating the true cylinder admittance matrix.  First consider $p$ as a function of polar coordinates such that Eq.~\eqref{8} becomes 
\begin{align}
\hspace{-30mm}v_r = \frac{-i}{\rho \omega (1+W^2)} \left( \left[ 1+ \frac{W^2}2\right] \diff{p}{r} + Wkp \cos(\theta - \phi) \right. \notag \\
 \left.- \frac{W^2}{2} \diff{p}{r} \cos 2(\theta - \phi) + \frac{W^2}{2r} \diff{p}{\theta} \sin 2(\theta - \phi) \right). \hspace{-30mm} \label{9}
\end{align}
Equation~\eqref{9} is relatively simple in $r,\theta$, as compared with the pressure in Eq.~\eqref{7}.  This suggests using the former in an exact sense combined with an approximation for $p$ in Eq.~\eqref{7} may lead to useful results. 

Consider the regime of  $W\ll 1$ for which a perturbation solution can be developed. 	Expanding $R$ and $\gamma$ of Eq.~\eqref{7} in the small parameter $W$ gives
%\beq{11}
%\ba R&= r \Big( 1 - \frac{W^2}4 +  \frac{W^2}4 \cos 2(\theta - \phi) 
%+\frac{W^4}{64} \big( 13 - 12 \cos 2(\theta - \phi) - \cos 4(\theta - \phi) \big) 
%+ \text{O}(W^6)\Big),
%\\ 
%\gamma &= (\theta - \phi)  - \frac{W^2}4 \sin 2(\theta - \phi)
%+\frac{W^4}{16} \big(2\sin 2(\theta - \phi)+\sin 4(\theta - \phi) \big) 
% + \text{O}(W^6).
%\ea	
%\eeq
\beq{11}
\ba R&= r\left\{1 - \frac{W^2}{4}\left[1 - \cos 2(\theta - \phi)\right] + \text{O}(W^4)\right\},
\\ 
\gamma &= (\theta - \phi)  - \frac{W^2}{4} \sin 2(\theta - \phi) + \text{O}(W^4).
\ea	
\eeq
Hence, the pressure and  the radial velocity may also be expanded to yield
\bse{121212}
\bal{12}
p =& \sum_n  B_n e^{i n \theta } 
\Big( J_n (kr)
-  \frac{W^2}4 \Big[krJ_n' (kr) \big( 1- \cos 2(\theta - \phi)\big) 
+ i n   J_n (kr) \sin 2(\theta - \phi) \Big] \Big)  
%\notag \\ &
 + \text{O}(W^4)~~\text{and}
\\
v_r 
 =& \frac {-i}{Z }  \sum_n B_n e^{i n \theta } 
\bigg( 
\big( 1- \frac{W^2}2\big) J_n' (kr)  +  \frac{W^2}4 \big( kr - \frac {n^2}{kr}\big)J_n (kr)
+ (W - W^3) J_n (kr)  \cos(\theta - \phi)  
\notag \\
 & %\qquad\qquad\qquad 
- \frac{W^2}4  \Big[\big[ 2J_n' (kr)   +\big( kr - \frac {n^2}{kr}\big)J_n (kr)\big] \cos 2(\theta - \phi)- in  \big[ \frac{2J_n (kr)}{kr} - J_n' (kr)\big]\sin 2(\theta - \phi)
\Big]\bigg) 
%\notag \\ & %\qquad\qquad\qquad 
 + \text{O}(W^4) .
\label{13}
\eal
\ese
Equations~\eqref{15} and \eqref{121212} imply that 
\beq{171}
P_m  =   \sum_n D_{mn} B_n, 
\ \ 
V_m  =   \sum_n E_{mn} B_n,
\eeq
from which  the admittance defined in \eqref{16} is  given by 
\beq{18}
Y_\text{cyl} = E D^{-1} . 
\eeq
Recalling that the anisotropic wave equation this analysis is based on is only valid up to $O(W^2)$, only terms up to $O(W^2)$ from Eq.~\eqref{121212} will be retained, which gives the admittance matrices $E$ and $D$ as  
\bse{2-2-2}
\bal{2-2}
D_{mn} =& \Big( J_n (ka) - ka \frac{W^2}4 J_n' (ka) \Big) \delta_{mn}
+ \frac{W^2}8\Big( ka J_n' (ka) \mp n J_n (ka)
\Big) e^{\mp i2 \phi} \delta_{m\, n\pm 2}
+ \text{O}(W^4), 
\\
E_{mn} =& \frac {i}{Z }  \bigg\{
\Big( J_n' (ka)    +  \frac{W^2}4 \Big[ \big( ka - \frac {n^2}{ka}\big)J_n (ka)
-2 J_n' (ka)\Big]  \Big) \delta_{mn} + \frac{W-W^3}{2}J_n (ka)
 e^{\mp i \phi} \delta_{m\, n\pm 1}
\notag \\
& -  \frac{W^2}8\Big[ \big(ka -\frac{n^2}{ka}\big) J_n (ka) +2 J_n' (ka) 
\mp n \big[ \frac{2}{ka} J_n (ka) -J_n' (ka) \big]
 \Big] e^{\mp i2 \phi} \delta_{m\, n\pm 2} \bigg\}
+   \text{O}(W^4) .
\label{3-}
\eal
\ese
These expressions may then be used to get a good approximation to the scattering.  Note that $\delta_{m\, n+1}$ is a diagonal string of ones below the main diagonal, $\delta_{m\, n-1}$ is above the main diagonal, and the symbols $\pm$ and $\mp$ should be treated as both the plus \emph{and} the minus cases (e.g., $(A\pm B)e^{i\mp c} \equiv (A +B)e^{-ic} + (A-B)e^{ic}$).

\subsubsection{First Order Approximation Revisited}

In order to compare the second order impedance approach with the first order approximation of Sec.~\ref{sec2B}, first write $D$ and $E$ of Eqs.~\eqref{2-2-2} as series in $W$, 
\beq{0=1}
\begin{aligned}
\  D  & = D^{(0)} {\color{white}+WE^{(1)}}\ + W^2D^{(2)} + \text{O}(W^4),
\\
E & = E^{(0)}+ WE^{(1)} + W^2E^{(2)} + \text{O}(W^3).
\end{aligned}
\eeq
It then follows from Eqs.~\eqref{18} and \eqref{2-2-2} that to $O(W)$ the impedance is 
\beq{0=2}
Y_\text{cyl}  = Y_\text{cyl}^{(0)}+ WY_\text{cyl}^{(1)} + \text{O}(W^2), 
\eeq
where $Y_\text{cyl}^{(0)}$ is a diagonal matrix and $Y_\text{cyl}^{(1)}$ is a bi-diagonal matrix
,
\bse{17}
\bal{17a}
 Y_{mn}^{(0) }&= \frac 1{Z_n} \delta_{m n} \ \  
\text{with } \  Z_n = -iZ \frac {J_n (ka)}{ J_n' (ka)}  , 
\\
 Y_{mn}^{(1) }&= \frac{i}{2Z} e^{\mp i \phi} \delta_{m\, n\pm 1}. 
\eal
\ese
Substitution from Eq.~\eqref{0=2} into Eq.~\eqref{b7} yields the  first order approximation of the  scattering matrix, 
\beq{22}
S = S^{(0)}+ WS^{(1)} + \text{O}(W^2), 
\eeq
where 
\beq{23}
 S^{(0)} =  \big(Y_\text{scat}+Y_\text{cyl}^{(0)} \big)^{-1}\big(Y_\text{inc}-Y_\text{cyl}^{(0)} \big),
\ \ \
 S^{(1)} =  -\big(Y_\text{scat}+Y_\text{cyl}^{(0)} \big)^{-1} Y_\text{cyl}^{(1)}
\big(I + S^{(0)} \big) .
\eeq
%$ S^{(1)} $  is the scattering correction due to the Born approximation.  
Substituting the two terms from Eq.~\eqref{22} into \eqref{1=-2} for  plane wave incidence $(F_n=i^n)$  it can be shown that 
$ S^{(0)}$  and $ S^{(1)}$ produce  the analytical expressions for the scattering amplitudes $A_m^{(0)}$ and $A_m^{(1)}$ given by  Eqs.~\eqref{eq:am0} and \eqref{3=9}, respectively. 
 The equivalence  has also been checked numerically. 
%O$(W)$ is identical to the perturbation solution derived in \cite{MBM2017}, specifically in their eqs.\ (14), (16)-(18).  This can be checked by assuming plane wave incidence (next) and comparing . 

 \subsection{Comparison of the First Order and Higher Order Approximations}

\begin{figure}
    \centering
    \includegraphics[scale=1.375]{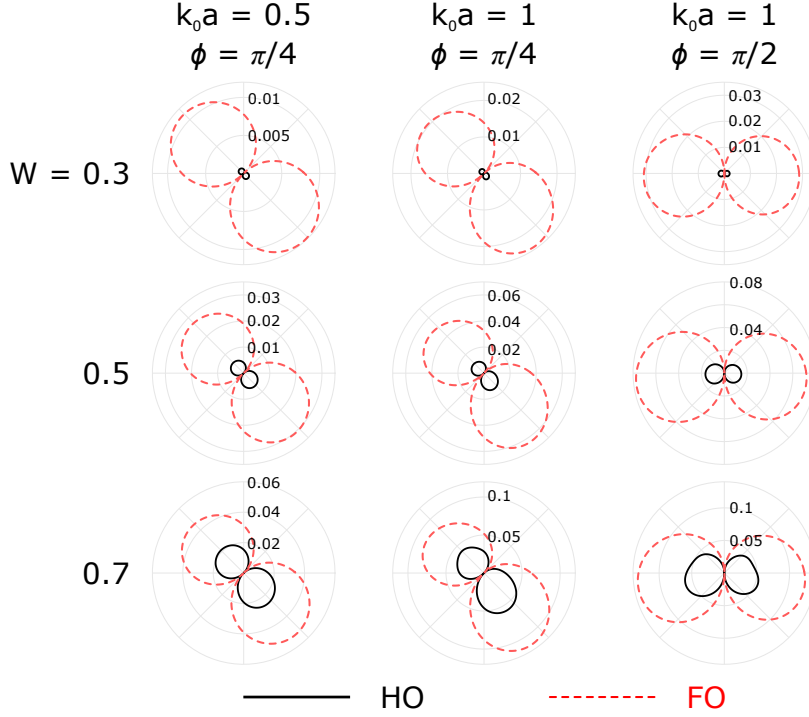}
\caption{\label{fig1}Comparison of the higher order (HO) approximation (solid) and the first order (FO) approximation (dashed) for values of $W$ from $W=0.3$ to $W=0.7$.  The polar plots show the values of the pressure boundary condition error $\Delta(\theta )$ of \eqref{5=3} for plane wave incidence from the left. Parameters common among all subfigures: $k/k_0=2/3$.}
\end{figure}

In order to compare the higher order (HO) approximation based on Eqs.~\eqref{b7}, \eqref{1=-2}, \eqref{18}, and \eqref{2-2-2} with the first order (FO)  approximation of \eqref{0-2} we consider how each satisfies the boundary conditions.  An exact solution will have perfect continuity of pressure and of radial particle velocity at the boundary $r=a$.  The approximate solutions will display discontinuities of these quantities to differing degrees.  Here we focus on the pressure condition, and define the angle-dependent parameter 
\beq{5=1}
\Delta p (\theta ) = p_\text{inc}(a,\theta ) + p_\text{scat}(a,\theta ) -  p_\text{cyl}(a,\theta ).
\eeq
Here  $p_\text{inc}$ and $ p_\text{scat}$ are defined by the incident and scattered fields in eq.~\eqref{1=1}.   The incident field is assumed to be  a plane wave of amplitude $p_0$.  The internal pressure $ p_\text{cyl}$ is defined by the exact series in Eq.~\eqref{7} with coefficients determined by Eq.~\eqref{171}.  That is, $C_m = e^{im\phi}\sum_{n} (D^{-1})_{mn} P_n$ where $P_n$ are the coefficients of the exterior pressure $p_\text{inc}(a) + p_\text{scat}(a) $, i.e. $P_n= F_n J_n(k_0a) + A_n H_n^{(1)}(k_0a)$ where $F_n = i^n$ and $A_n$ are determined by with the FO or the HO approximation.  The matrix $D$ is defined by Eq.~\eqref{2-2} for the HO approximation and is $D_{mn} = J_n(ka) \delta _{mn}$ for the FO approximation.
Thus, 
\beq{5=2}
\Delta p  (\theta )=  \sum_n  \Big\{ \big( F_n J_n(k_0a) + A_n H_n^{(1)}(k_0a)\big) \, 
e^{i n \theta } 
- C_n J_n(kR|_{r=a}) e^{i n \gamma } \Big\}
.
\eeq
In the following, we present comparisons of the quantity 
\begin{equation}\label{5=3}
	\Delta(\theta) = \left| \frac{\Delta p(\theta) }{p_0} \right|. 
\end{equation}
Examples for various values of relevant parameters are shown in Figure \ref{fig1}.  The parameters in Figure \ref{fig1} cover a wide range of those physically admissible in terms of frequency and impedance.  In particular, we note that the HO approximation show smaller discontinuity in the pressure over a wide range of the perturbation parameter $W$, up to $0.7$.

 \subsubsection{Scattered far-field}
\label{Far field}
\begin{figure}
    \centering
    \includegraphics[scale=1.375]{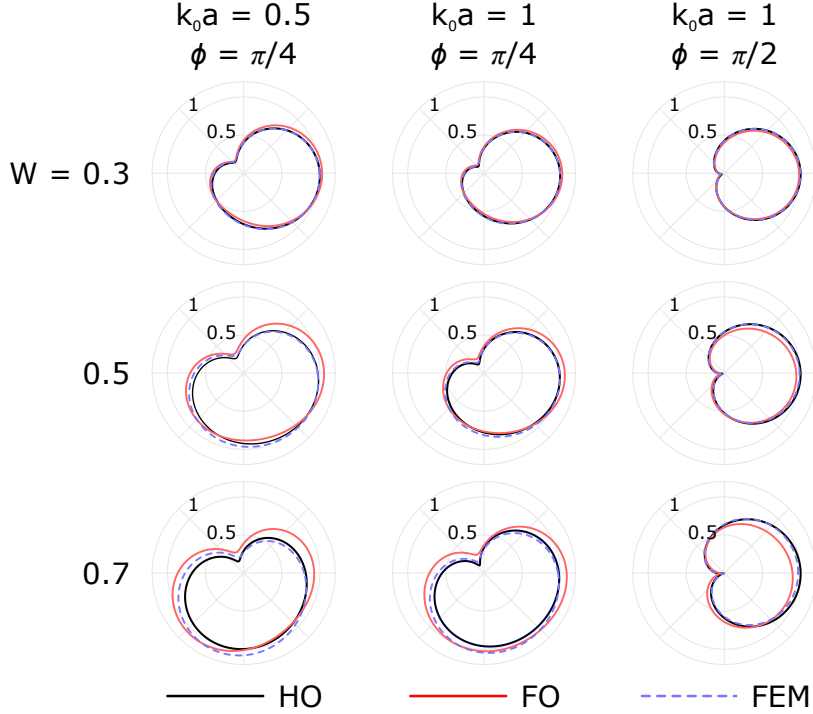}
    \caption{Comparison of the far-field scattering amplitudes using the higher order (HO)  approximation and the first order (FO) approximation for plane wave incidence from the left.  Predictions using a finite-element method (FEM) that are described in Sec.~\ref{sec:numeric} are also shown. Parameters common among all subfigures: $k/k_0=  2/3$.}
\label{fig2}
\end{figure}

The comparisons of Figure \ref{fig1} provide confidence that the HO approximation
provides more accurate  estimates of the scattered pressure for plane wave incidence.  Based on this, we show in Figure \ref{fig2} the far-field amplitude for different values of the parameters $Z$, $\phi$, $k_0a$ and for values of $W$ as large as $0.7$.    These plots indicate that the first order Born approximation good  for values of $W$ less than $0.5$.  For larger values the HO approximation indicates different scattering patterns and amplitudes, particularly in some scattering directions.  

%%%%%%%%%%%%%%%%%%%%%%%%%%%%%%%%%%%%%%%%%%%%%%%%%%%%%%%%%%%%%%%%%%%%%%%%%%%%%%%%%%%%%%%%

\section{Finite Element Analysis}\label{sec:numeric}
A model based on the finite element method (FEM) has been derived to assess the accuracy of the approximations in Sec.~\ref{sec2}. This model provides a solution of the full scattering problem whose error is independent of the magnitude of the Willis coupling vector. Instead, the sources of error are the familiar inaccuracies associated with FEM, including the discretization of the geometry into a triangular mesh, and the projection of the scattered field onto a finite set of basis functions. However, the overall error is bounded by the size of the mesh elements, which decreases as the mesh is refined.\cite{gockenbach2006}

The geometry of the problem under consideration is shown in Fig.~\ref{schematic}. As with the analytical  solutions presented in Secs.~\ref{sec2B} and \ref{sec3b}, the FEM will only consider the two-dimensional case since all fields are assumed constant along the $z$-axis.
The resulting far field calculation of the pressure field will take into account the invariance of the solution along the axis of the cylinder by utilizing the scattered field expansion in Eq.~\eqref{eq:psca}.  The implementation of FEM requires a variational formulation for the wave equation, often called the weak formulation, that accounts for Willis coupling in the constitutive equations. The derivation of the weak formulation is provided below, followed by the projection of the computed scattered field solution to the far field.

\subsection{Weak Form}
\begin{figure}
	\centering
	\includegraphics[width=2.8in]{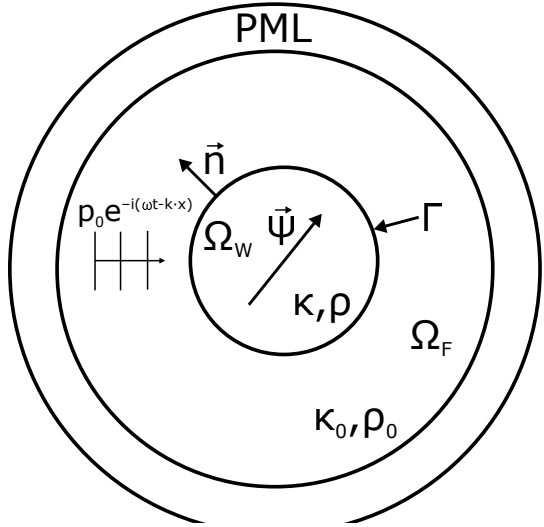}
	\caption{\label{fig:computationalDomain}FEM computational domain, where $\Omega_\text{W}$ is the Willis cylinder, $\Omega_\text{F}$ is the exterior fluid domain, and PML is the perfectly matched layer used to truncate the computational domain.}
\end{figure}

Let $\Omega_\text{W}$, $\Omega_\text{F}$, and $\Gamma$ represent the domains of the Willis fluid, the exterior fluid, and the boundary shared between the two domains, respectively; see Fig.~\ref{fig:computationalDomain}. 
The weak form for the acoustic pressure in the Willis domain is derived by multiplying the time-harmonic equation $\nabla \cdot \vec{v} + i\omega \varepsilon = 0$ by a test function $\phi_\text{cyl}$ and integrating over the Willis domain to yield the integral equation
\begin{equation}
\int \limits_{\Omega_\text{W}} \left(\nabla \cdot \vec{v}\right) \phi_\text{cyl} \, d\Omega_\text{W} + i \omega \int \limits_{\Omega_\text{W}} \varepsilon \phi_\text{cyl} \, d\Omega_\text{W} = 0.
\end{equation}
Utilizing Green's identity on the first integral gives the equation
\begin{equation}
\label{eq:GreensEq}
-\frac{i}{\omega}\int \limits_{\Omega_\text{W}} \vec{v} \cdot \nabla \phi_\text{cyl} \, d\Omega_\text{W} - \int \limits_{\Omega_\text{W}} \varepsilon \phi_\text{cyl} \, d\Omega_\text{W} + \frac{i}{\omega} \int \limits_\Gamma \phi_\text{cyl} \left(\vec{v} \cdot \vec{n} \right) \, d\Gamma = 0,
\end{equation}
where it is assumed that the boundary of the Willis medium completely shares a boundary with the exterior fluid. 
The relationship for volume strain and velocity fields provided in Eqns.~\eqref{5a}-\eqref{simpVel}, are substituted in Eq.~\eqref{eq:GreensEq} to yield the weak form for the acoustic pressure in a Willis fluid
\begin{multline}
\label{eq:WillisWeakForm}
\frac{-1}{\omega^2\rho (1+W^2)} \int \limits_{\Omega_\text{W}} \biggl[(1+W^2)\nabla p_\text{cyl} \cdot \nabla \phi_\text{cyl} - k^2p_\text{cyl}\phi_\text{cyl} +k p_\text{cyl} \left(\vec{W} \cdot \nabla \phi_\text{cyl}\right) + k \phi_\text{cyl} \left(\vec{W} \cdot \nabla p_\text{cyl}\right) \\
- \left(\vec{W}\cdot \nabla p_\text{cyl}\right)\left(\vec{W}\cdot \nabla \phi_\text{cyl}\right)\biggr] \, d\Omega_\text{W} + \frac{i}{\omega} \int \limits_\Gamma \phi_\text{cyl} \left(\vec{v}\cdot \vec{n}\right) \, d\Gamma = 0.
\end{multline}
Similarly, the weak form for the scattered pressure in the exterior fluid, $p_\text{scat}$, may be found to be\cite{ihlenburg2006}
\begin{multline}
\label{eq:WeakFormFluid}
\frac{-1}{\rho_0 \omega^2} \int \limits_{\Omega_\text{F}} \left(\nabla p_\text{scat} \cdot \nabla \phi_\text{scat} - k_0^2 p_\text{scat} \phi_\text{scat}\right)\ d\Omega_\text{F} - \frac{i}{\omega}\int \limits_\Gamma \left(\vec v \cdot \vec n\right) \phi_\text{scat} \ d\Gamma \\
=  \frac{1}{\rho_0\omega^2}\int \limits_\Gamma \phi_\text{scat} \left(\nabla p_\text{inc} \cdot \vec n\right) \ d\Gamma,
\end{multline}
where $\phi_\text{scat}$ is the test function of the scattered pressure field in the exterior fluid. 
Equations \eqref{eq:WillisWeakForm} and \eqref{eq:WeakFormFluid} are combined to yield the total integral equation for the coupled fields $(p_\text{cyl},p_\text{scat})$
\begin{equation}
\label{eq:WeakFormInt}
\mathcal{I}_{\Omega_\text{W}} + \mathcal{I}_{\Omega_\text{F}} = \mathcal{I}_\text{inc},
\end{equation}
where $\mathcal{I}_{\Omega_\text{W}}$ is the volume integral in Eq.~\eqref{eq:WillisWeakForm}, $\mathcal{I}_{\Omega_\text{F}}$ is the volume integral in Eq.~\eqref{eq:WeakFormFluid}, and $\mathcal{I}_\text{inc}$ is the surface integral on the right hand side of Eq.~\eqref{eq:WeakFormFluid}. 
The surface integrals in Eqns.~\eqref{eq:WillisWeakForm} and \eqref{eq:WeakFormFluid} are used to describe the continuity of normal particle velocity at the interface, as described below.
Galerkin's method is used to numerically solve Eq.~\eqref{eq:WeakFormInt} by seeking approximations to $p_\text{cyl}$ and $p_\text{scat}$ that are written as a linear combination of basis functions, which are chosen to be piecewise quadratic Lagrange polynomials.\cite{reddy1993} 
The continuity of pressure on the interface, $p_\text{cyl} = p_\text{inc} + p_\text{scat}$, is directly enforced on the basis functions. 
The weak form is made symmetric by choosing test functions that are represented with the same basis functions as the unknown dependent variables. 
Given this choice in test functions supplemented with the fact that the incident wave is known, the continuity of normal particle velocity is naturally enforced
\begin{equation}
\frac{i}{\omega} \int \limits_\Gamma \phi_\text{cyl} \left(\vec{v}\cdot \vec{n}\right) \, d\Gamma - \frac{i}{\omega}\int \limits_\Gamma \phi_\text{scat} \left(\vec v \cdot \vec n\right) \ d\Gamma = 0,
\end{equation}
and therefore does not appear in Eq.~\eqref{eq:WeakFormInt}.
Equation \eqref{eq:WeakFormInt} is solved using the finite element software package COMSOL Multiphysics using the weak form PDE module and making use of built-in perfectly matched layers (PML) to truncate the computational domain (see Fig.~\ref{fig:computationalDomain}) and satisfy the Sommerfeld radiation condition.\cite{ihlenburg2006,2017}  A sufficiently fine mesh of the computational domain was used to obtain convergence.  It is also worthwhile to note that the exterior fluid can be modeled with a boundary integral method instead of Eq.~\eqref{eq:WeakFormFluid}.  A hybrid finite-element/boundary-element method may be obtained using similar techniques in acoustic fluid-structure interaction problems.\cite{demkowicz1996}

\subsection{Far field calculation}
The far field solution is found by numerically calculating the scattered field coefficients in Eq.~\eqref{eq:psca}, which can then be compared to the approximate solutions found using the methods developed in Sec.~\ref{sec2}. The resulting $p_\text{scat}$ from FEM at a radius $b$ is expanded into outward-propagating cylindrical waves
\begin{equation}
p_\text{scat}(r=b,\theta) = \sum_{-\infty}^\infty A_m H_m^{(1)} (k_0 b) e^{im\theta},
\end{equation}
where $b$ is chosen to be a sufficient distance away from the cylinder surface such that the evanescent waves are attenuated. Numerical studies show that a radius of one wavelength from the cylinder surface is sufficient. Orthogonality in $\theta$ is used to uniquely determine the scattered field coefficients using a Fourier integral
\begin{equation}
A_m = \frac{1}{2 \pi H_m^{(1)}(k_0 b)} \int_{-\pi}^{\pi} p_\text{scat}(r=b,\theta) e^{-im\theta}\ d\theta.
\end{equation} 
The above integral is a Fourier transform which may be numerically approximated using optimized algorithms such as the Fast Fourier Transform (FFT).

The directivity pattern for the FEM prediction is shown in Fig.~\ref{fig2} along with the analytical predictions.  As may be seen in Fig.~\ref{fig2}, the FEM and analytical predictions are nearly identical for $W=0.3$, and the behavior of each prediction is qualitatively the same for all cases shown.  As $W$ increases the FEM prediction is consistently closer to the HO prediction than to the FO prediction.  The similarity of the HO and FEM predictions despite their different approaches suggests that the results are accurate.

\begin{figure}
	\centering
	\includegraphics[width=3.315in]{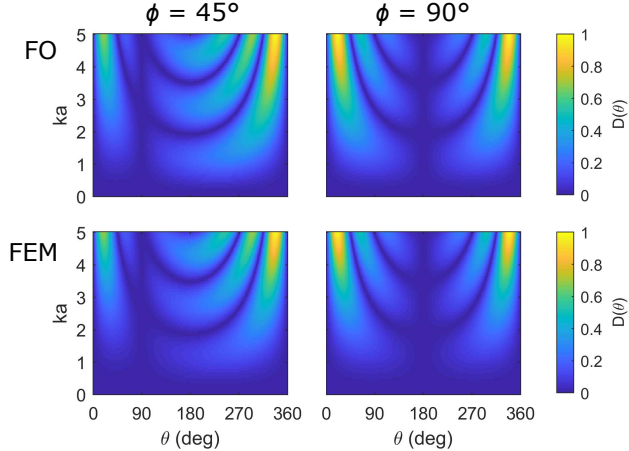}
%	\begin{tabular}{lccc}
%	\multicolumn{2}{r}{$\phi=45^\circ$}\hspace{11mm} & & $\phi = 90^\circ$\hspace{15mm}
%	\\ \hline
%	\raisebox{6\height}{FO}
%	&\rule{0.5pt}{40mm}\includegraphics[trim={0 9mm 4mm 0},clip]{FO_SurfacePlot_45deg}
%	& ~ &\includegraphics[trim={8mm 9mm 4mm 0},clip]{FO_SurfacePlot_90deg}
%	\includegraphics[trim={36mm 9mm 0 0},clip]{SurfacePlot_Colorbar}
%	\\
%	\raisebox{10\height}{FEM}\hspace{1mm}
%	&\raisebox{5mm}{\rule{0.5pt}{42mm}}\includegraphics[trim={0 0 4mm 0},clip]{FEM_SurfacePlot_45deg}
%	& &\includegraphics[trim={8mm 0 4mm 0},clip]{FEM_SurfacePlot_90deg}
%	\includegraphics[trim={36mm 0 0 0},clip]{SurfacePlot_Colorbar}
%	\end{tabular}
	\caption{\label{fig:surfaceplots}Scattered directivity of a plane wave incident upon a Willis-coupled cylinder.  The directivity is shown as a function of the wavenumber times the cylinder radius $ka$ and of scattered angle $\theta$.  Directivities are shown using the first-order (FO) approximation and a finite-element method (FEM) and for $\phi=45^\circ$ and $\phi=90^\circ$, where $\phi$ represents the orientation of the Willis coupling vector.}
\end{figure}

Plots of the directivity patterns of the FO and FEM predictions as a function of $ka$ and $\theta$ are shown in Fig.~\ref{fig:surfaceplots} for $\phi=45^\circ$ and $\phi=90^\circ$.  For these plots $W=0.1$ and $k_0a=ka$.  As may be seen in the figure, there is no discernible difference between the FO and FEM predictions, suggesting that the FO approximation is sufficient to provide accurate predictions for these conditions.

\section{Conclusions}\label{sec:conclusions}

Two types of approximations have been derived for small values of the the non-dimensional Willis coupling $W$.  Numerical results compare the first order O$(W)$ and higher order O$(W^3)$  approximations  in terms of how well they satisfy the pressure boundary condition.  As expected the HO approximation shows less error.  Comparisons of the scattered far-fields indicate that the O$(W)$  approximation does not differ much from the O$(W^3)$  approximation for $W \le 0.7$.  This suggests that the simpler FO approximation may as well be used, especially for smaller values of $W$.

Furthermore, a finite-element method for predicting the far-field scattering pattern that is not limited to small $W$ or cylindrical geometry of the scatterer has been developed and implemented.  This additional method was compared with the first and higher order analytical approximations and good agreement is found for $W\le0.7$.  The correlation of the numerical and analytical predictions provides support for the both solutions and suggests that Willis coupling does indeed modify the far-field scattering pattern in measurable ways.

\appendix   %%%%%%%%%%%%%%%%%%%%%%%%%%%%%%%%%%%%%%%%%%%%%%%%%%%%%%%%%%%%%%%%%%%%%%%%%%%%%

\section{First Order Particle Velocity Boundary Condition}\label{appendix:velocity}

The first order approximation to the  normal component of the incident, scattered, and interior particle velocities at the surface of the cylinder are given by
\begin{align}
	\hat r\cdot{\vec v}_\inc|_a &= \left.-\frac{1}{k_0Z_0}\diff{p_\inc}{r}\right|_{r=a} = \frac{-p_0}{Z_0}\sum_{m=-\infty}^\infty i^mJ'_m(k_0a)e^{im\theta}, \\
	\hat r\cdot{\vec v}_\sca|_a &= \left.-\frac{1}{k_0Z_0}\diff{p_\sca}{r}\right|_{r=a} = \frac{-p_0}{Z_0}\sum_{m=-\infty}^\infty A_mH^{(1)'}_m(k_0a)e^{im\theta},
\end{align}
and
\begin{align}
	\hat r\cdot{\vec v}_\cyl|_a &= \left.-\frac{1}{kZ}\diff{p_\cyl}{r}\right|_{r=a} - \left.i\frac{\hat r\cdot\vec W}{Z}p_\cyl\right|_{r=a} %\notag \\	& 
	= -\frac{p_0}{Z}\sum_{m=-\infty}^\infty \left[ B_mJ'_m(ka) + i\hat r\cdot\vec W B_mJ_m(ka) \right] e^{im\theta} .
\end{align}
 In order to apply orthogonality, all of the $\theta$ dependence should be represented by the $e^{im\theta}$ term, which is not the case in the form due to the presence of $\hat r$.  Note that
\begin{equation}
	\hat r\cdot\vec W =  W\cos(\theta-\phi) = \frac{W}{2}\left[ e^{i\theta}e^{-i\phi} + e^{-i\theta}e^{i\phi} \right].
\end{equation}
Then we may write
\begin{align}
	\sum_{m=-\infty}^\infty \hat r\cdot\vec W B_mJ_m(ka)e^{im\theta} 
	&= \frac{W}{2}\sum_{m=-\infty}^\infty \left[ e^{i\theta}e^{-i\phi} + e^{-i\theta}e^{i\phi} \right] B_mJ_m(ka)e^{im\theta} \\
	%&= \frac{W}{2}\sum_{m=-\infty}^\infty \left[ e^{i(m+1)\theta}e^{-i\phi}B_mJ_m(ka) + e^{i(m-1)\theta}e^{i\phi}B_mJ_m(ka) \right] \\
	&= \frac{W}{2}\sum_{m=-\infty}^\infty \left[ e^{-i\phi}B_{m-1}J_{m-1}(ka) + e^{i\phi}B_{m+1}J_{m+1}(ka) \right]e^{im\theta}.
\end{align}
Now, applying orthogonality results in the condition \eqref{eq:Vbc}.

\section*{Data accessibility}

No data has been generated associated with this paper.

\section*{Competing interests}

We declare we have no competing interests.

\section*{Authors' contributions}

M.B.M. derived the first-order approximation and A.N.N. derived the higher-order approximation.  B.M.G. developed and implemented the finite-element analysis.  M.R.H. provided valuable insight which guided the derivations given.  All authors gave final approval for publication.

\section*{Acknowledgements}

The work for this paper was performed independently by the authors.

\section*{Funding Statement}
This work was supported in part by ONR through MURI Grant No. N00014-13-1-0631 and YIP Grant No. N0014-18-1-2335, and in part by the U.S Army Engineer Research and Devlopment Center (ERDC), Geospatial Research and Engineering business area.  Permission to publish was granted by the Director, Cold Regions Research and Engineering Laboratory.

\section*{Ethics statement}

No research on living or dead organisms was performed during the preparation of this paper.

%%%%%%%%%%%%%%%%%%%%%%%%%%%%%%%%%%%%%%%%%%%%%%%%%%%%%%%%%%%%%%%%%%%%%%%%%%%%%%%%%%%%%%%%%%%%%%%%%%
%\bibliography{scattering2}

\end{document}